\documentclass[twocolumn]{revtex4}

\usepackage{graphicx}
\usepackage{dcolumn}
\usepackage{amsmath}

\begin{document}

\title{Kondo effect in Dirac systems
}

\author{Takashi YANAGISAWA}

\affiliation{Electronics and Photonics Research Institute,
National Institute of Advanced Industrial Science and Technology (AIST),
Tsukuba Central 2, 1-1-1 Umezono, Tsukuba 305-8568, Japan
}


\begin{abstract}
We investigate the Kondo effect in Dirac systems, 
where Dirac electrons interact with the localized spin via the s-d
exchange coupling.
The Dirac electron in solid state has the linear dispersion and is
described typically
by the Hamiltonian such as $H_k= v{\bf k}\cdot {\sigma}$ for the 
wave number ${\bf k}$ 
where $\sigma_j$ are Pauli matrices.
We derived the formula of 
the Kondo temperature $T_{\rm K}$ by means of 
the Green's function theory for small $J$.
The $T_{\rm K}$ is determined from a singularity of Green's functions
in the form $T_{\rm K}\simeq \bar{D}\exp(-{\rm const.}/\rho |J|)$ when
the exchange coupling $|J|$ is small 
where $\bar{D}=D/\sqrt{1+D^2/(2\mu)^2}$ for a cutoff $D$ and $\rho$
is the density of states at the Fermi surface. 
When $|\mu|\ll D$, $T_{\rm K}$ is proportional to $|\mu|$:
$T_{\rm K}\simeq |\mu|\exp(-{\rm const.}/\rho |J|)$.
The Kondo screening will, however, disappear when the Fermi surface 
shrinks to a point called the Dirac point, that is,  
$T_{\rm K}$ vanishes when the chemical potential $\mu$ is just
at the Dirac point.
The resistivity and the specific heat exhibit a log-$T$ singularity in the range
$T_{\rm K} < T\ll |\mu|/k_{\rm B}$.
Instead, for $T\sim O(|\mu|)$ or $T>|\mu|$, they never show log-$T$.
\end{abstract}
 

\maketitle

\section{Introduction}

Recently, the Dirac electron in solid state has been investigated 
intensively because the Dirac cones in metals or semimetals have been 
realized in solids\cite{nov05, zha05, hsi08, mit13, ber12}.
In several materials such as Graphene\cite{cas09,das11, kot12, kan12, jaf13} 
as well as Bismuth compounds, there
appear the conduction bands with the linear dispersion that are described
by Dirac Hamiltonian.
The Dirac cone in the band structure appears in the surface states of 
topological insulators\cite{has10,qi11}.
Three-dimensional Dirac semimetals have also been realized in
Na$_3$Bi\cite{liu14} and Cd$_3$As$_2$\cite{neu14}. 

The Kondo effect, that occurs as the result of the exchange interaction
between  dilute magnetic impurities
and conduction electrons, is one of the most important phenomena in
solid state physics\cite{kon64,kondo}.
It is interesting to examine how the Kondo effect emerges in Dirac systems. 
The linear dispersion of Dirac electrons would affect the
characteristic features of the Kondo effect.
It is not even trivial whether the Kondo effect indeed appears in Dirac systems.
The purpose of this work is to show unique and interesting properties
of the Kondo effect in these systems.

The s-d model in Dirac systems is closely related to the pseudogap 
model of the Kondo problem where magnetic impurities couple to conductive 
fermions with a density of states 
$\rho(\omega)\propto |\omega|^r$ ($0<r$)\cite{gon98, voj06, fri04}.
In the pseudogap model, the density of states vanishes at the Fermi level.
This feature is common to the Kondo problem in Dirac systems when
the Fermi surface is point like, namely, the chemical potential is just at the
Fermi point $\mu=0$.
It has been suggested that there is a phase transition between the 
local-moment phase and the strong-coupling singlet phase in the
pseudogap Kondo and Anderson models\cite{voj06}.
A Dirac system is also interesting from the viewpoint of topology
because the index theorem has been proved for Dirac 
operators\cite{roe88, mel93}

In this paper we investigate the s-d Hamiltonian in a Dirac system by means 
of the Green's function theory and evaluate the Kondo temperature $T_{\rm K}$.
We employ the decoupling procedure to obtain a closed solution for a set
of equations for Green's functions.  
Although the decoupling procedure is valid only for small $J$ in the
region, $T>T_{\rm K}$, this method is useful to derive the Kondo temperature.   
It turns out that $T_{\rm K}$ is crucially dependent on the chemical potential $\mu$.
$T_{\rm K}$ vanishes for $\mu=0$ at the Dirac point in our method.
This shows the absence of Kondo screenings at $\mu=0$ and
suggests that there may be a transition at some critical value of 
$J=J_{\rm c}$ from the  local-moment phase
to the singlet-formation phase at $\mu=0$.
In Dirac systems $T_{\rm K}$ is finite and is given by the
formula being similar to that in the conventional Kondo effect:
$T_{\rm K} \propto \bar{D}\exp(-{\rm const.}/\rho |J|)$  when $\rho |J|$ is small
where $\rho$ is the density 
of states and $J$ is the exchange coupling constant.
When $\rho |J|$ is large of order 1, $T_{\rm K}$ is given by an algebraic function of
$\rho|J|$ like
$T_{\rm K}\propto |\mu|Q(\rho|J|)$ for a some function $Q(x)$.

The paper is organized as follows.  In section II we show the Hamiltonian.
In section III Green's functions are defined and a set of equations for them are
derived.  In section IV the Kondo temperature is derived.
In sections V and VI, we evaluate the electric resistivity and specific heat, respectively,
and discuss their log-$T$ singularity.
The last sect ion is devoted to a summary.

\section{Hamiltonian}

We consider a massless Dirac Hamiltonian (Weyl Hamiltonian):
\begin{equation}
H_D= \sum_{{\bf k}}\psi_{{\bf k}}^{\dag}\left( v_xk_x\sigma_x
+v_yk_y\sigma_y+v_zk_z\sigma_z \right)\psi_{{\bf k}},
\end{equation}
where
\begin{eqnarray}
\psi_{{\bf k}}=\left(
\begin{array}{c}
c_{{\bf k}\uparrow} \\
c_{{\bf k}\downarrow} \\
\end{array}
\right)
\end{eqnarray}
with the annihilation and creation operators
$c_{{{\bf k}}\sigma}$ and $c_{{{\bf k}}\sigma}^{\dag}$, respectively.
Here $v_x$, $v_y$ and $v_z$ are velocities of conduction electrons.
$\sigma_x$, $\sigma_y$ and $\sigma_z$ are Pauli matrices.
In our model the Dirac electrons, described by this Hamiltonian, interact
with the localized spin.
The total Hamiltonian is written as  $H=H_0+H_{sd}$ where
\begin{align}
H_0&= \sum_{{\bf k}}[ (v_xk_x-iv_yk_y)c_{{\bf k}\uparrow}^{\dag}
c_{{\bf k}\downarrow}
+(v_xk_x+iv_yk_y)c_{{\bf k}\downarrow}^{\dag}c_{{\bf k}\uparrow}\nonumber\\
&~ + v_zk_z(c_{{\bf k}\uparrow}^{\dag}c_{{\bf k}\uparrow}
-c_{{\bf k}\downarrow}^{\dag}c_{{\bf k}\downarrow})
-\mu(c_{{\bf k}\uparrow}^{\dag}c_{{\bf k}\uparrow}
+c_{{\bf k}\downarrow}^{\dag}c_{{\bf k}\downarrow})],
\\
H_{sd}&= -\frac{J}{2}\frac{1}{N}\sum_{{\bf k}{\bf k}'}[
S_z(c_{{\bf k}\uparrow}^{\dag}c_{{\bf k}'\uparrow}-
c_{{\bf k}\downarrow}^{\dag}c_{{\bf k}'\downarrow})
+S_{+}c_{{\bf k}\downarrow}^{\dag}c_{{\bf k}'\uparrow}\nonumber\\
&~ +S_{-}c_{{\bf k}\uparrow}^{\dag}c_{{\bf k}'\downarrow}].
\end{align}
$N$ is the number of sites and we have included the chemical potential $\mu$.
$S_{+}$, $S_{-}$ and $S_z$ denote the operators of the localized spin.
The term $H_{sd}$ indicates the s-d interaction between the conduction electrons
and the localized spin, with the coupling constant $J$\cite{kon64,kondo}.
$J$ is negative, as adopted in this paper, for the antiferromagnetic interaction.
The Dirac Hamiltonian resembles the s-d model with the 
spin-orbit coupling of Rashba type\cite{yan12}.
We use the Green's function method, following Ref.\cite{yan12} to evaluate 
the Kondo temperature and
its related properties.

\section{Green's Functions}

We define thermal Green's functions\cite{zub60,agd}:
\begin{align}
G_{{\bf k}{\bf k}'\sigma}(\tau)&=-\langle T_{\tau}c_{{\bf k}\sigma}(\tau)
c_{{\bf k}'\sigma}^{\dag}(0)\rangle,\\
F_{{{\bf k}}{{\bf k}}'}(\tau)&=-\langle T_{\tau}c_{{{\bf k}}\downarrow}(\tau)
c_{{{\bf k}}'\uparrow}^{\dag}(0)\rangle,\\
\langle\langle S_zc_{{{\bf k}}\uparrow};c_{{{\bf k}}'\uparrow}^{\dag}
\rangle\rangle_{\tau}
&= -\langle T_{\tau}S_z c_{{{\bf k}}\uparrow}(\tau)c_{{{\bf k}}'\uparrow}^{\dag}
(0)\rangle,\\
\langle\langle S_{-}c_{{{\bf k}}\downarrow};c_{{{\bf k}}'\uparrow}^{\dag}
\rangle\rangle_{\tau}
&= -\langle T_{\tau}S_{-}c_{{{\bf k}}\downarrow}(\tau)c_{{{\bf k}}'\uparrow}^{\dag}(0)
\rangle,\\
\langle\langle S_zc_{{{\bf k}}\uparrow};c_{{{\bf k}}'\uparrow}^{\dag}
\rangle\rangle_{\tau}
&= -\langle T_{\tau}S_z c_{{{\bf k}}\uparrow}(\tau)c_{{{\bf k}}'\uparrow}^{\dag}(0)
\rangle,\\
\langle\langle S_-c_{{{\bf k}}\downarrow};c_{{{\bf k}}'\uparrow}^{\dag}
\rangle\rangle_{\tau}
&= -\langle T_{\tau}S_- c_{{{\bf k}}\downarrow}(\tau)c_{{{\bf k}}'\uparrow}^{\dag}(0)
\rangle.
\end{align}
Here $T_{\tau}$ is the time ordering operator.
The Fourier transforms are defined as usual:
\begin{align}
G_{{{\bf k}}{{\bf k}}'\sigma}(\tau)&= \frac{1}{\beta}\sum_n e^{-i\omega_n\tau}
G_{{{\bf k}}{{\bf k}}'\sigma}(i\omega_n),\\
F_{{{\bf k}}{{\bf k}}'}(\tau)&= \frac{1}{\beta}\sum_n e^{-i\omega_n\tau}
F_{{{\bf k}}{{\bf k}}'}(i\omega_n),\\
\langle\langle S_zc_{{{\bf k}}\uparrow};c_{{{\bf k}}'\uparrow}^{\dag}
\rangle\rangle_{\tau}&=
\frac{1}{\beta}\sum_n e^{-i\omega_n\tau}
\langle\langle S_zc_{{{\bf k}}\uparrow};c_{{{\bf k}}'\uparrow}^{\dag}
\rangle\rangle_{i\omega_n},\\
\langle\langle S_-c_{{{\bf k}}\downarrow};c_{{{\bf k}}'\uparrow}^{\dag}
\rangle\rangle_{\tau}&=
\frac{1}{\beta}\sum_n e^{-i\omega_n\tau}
\langle\langle S_-c_{{{\bf k}}\downarrow};c_{{{\bf k}}'\uparrow}^{\dag}
\rangle\rangle_{i\omega_n}.
\end{align}
We also use the following Green's function,
\begin{align}
\Gamma_{{{\bf k}}{{\bf k}}'\uparrow}(\tau)&=\frac{1}{\beta}\sum_n e^{-i\omega_n}
\Gamma_{{{\bf k}}{{\bf k}}'\uparrow}(i\omega_n)\nonumber\\
&= \langle\langle S_zc_{{{\bf k}}\uparrow};c_{{{\bf k}}'\uparrow}^{\dag}
\rangle\rangle_{\tau}
+ \langle\langle S_-c_{{{\bf k}}\downarrow};c_{{{\bf k}}'\uparrow}^{\dag}
\rangle\rangle_{\tau}.
\end{align}

We start from the commutation relations:
\begin{align}
\left[H_0,c_{{\bf k}\uparrow}\right]&= -(v_xk_x-iv_yk_y)c_{{\bf k}\downarrow}
+(-v_zk_z+\mu) c_{{\bf k}\uparrow},\\
\left[H_0,c_{{\bf k}\downarrow}\right]&= -(v_xk_x+iv_yk_y)c_{{\bf k}\uparrow}
+(v_zk_z+\mu) c_{{\bf k}\downarrow},\\
\left[H_{sd},c_{{\bf k}\uparrow}\right]&= -\frac{J}{2N}\sum_{{\bf k}'}
(-S_z c_{{\bf k}'\uparrow}-S_{-}c_{{\bf k}'\downarrow} ),\\
\left[H_{sd},c_{{\bf k}\downarrow}\right]&= -\frac{J}{2N}\sum_{{\bf k}'}
(S_z c_{{\bf k}'\downarrow}-S_{+}c_{{\bf k}'\uparrow} ).
\end{align}
The equations of motion for $G_{{{\bf k}}{{\bf k}}'\uparrow}(\tau)$ and
$F_{{{\bf k}}{{\bf k}}'}$ are
\begin{align}
\frac{\partial}{\partial\tau}G_{{{\bf k}}{{\bf k}}'\uparrow}(\tau)&=
-\delta(\tau)\delta_{{{\bf k}}{{\bf k}}'}
+(-v_zk_z+\mu) G_{{{\bf k}}{{\bf k}}'\uparrow}(\tau)\nonumber\\
&~- (v_xk_x-iv_yk_y)F_{{{\bf k}}{{\bf k}}'}(\tau)\nonumber\\
&~ +\frac{J}{2N}\sum_{{{\bf q}}}[
\langle\langle S_zc_{{{\bf q}}\uparrow};c_{{{\bf k}}'\uparrow}^{\dag}
\rangle\rangle_{\tau}
+ \langle\langle S_-c_{q\downarrow};c_{{{\bf k}}'\uparrow}^{\dag}
\rangle\rangle_{\tau} ],
\nonumber\\
\end{align}
\begin{align}
\frac{\partial}{\partial\tau}F_{{{\bf k}}{{\bf k}}'}(\tau)&= (v_zk_z+\mu)
F_{{{\bf k}}{{\bf k}}'}
-(v_xk_x+iv_yk_y)G_{{{\bf k}}{{\bf k}}'\uparrow}(\tau)\nonumber\\
&~ - \frac{J}{2N}\sum_q [
\langle\langle S_zc_{{{\bf q}}\downarrow};c_{{{\bf k}}'\uparrow}^{\dag}
\rangle\rangle_{\tau}
- \langle\langle S_+c_{{{\bf q}}\uparrow};c_{{{\bf k}}'\uparrow}^{\dag}
\rangle\rangle_{\tau} ].\nonumber\\
\end{align}
Then the equation for $G_{{\bf k}{\bf k}'\uparrow}$ reads
\begin{align}
&(i\omega_n+\mu-v_zk_z)G_{{\bf k}{\bf k}'\uparrow}(i\omega_n) =
\delta_{{\bf k}{\bf k}'}\nonumber\\
& +(v_xk_x-iv_yk_y)F_{{\bf k}{\bf k}'}(i\omega_n)
 -\frac{J}{2N}\sum_{{\bf q}}\Gamma_{{\bf q}{\bf k}'\uparrow}(i\omega_n).
\end{align}
The equations for 
$\langle\langle S_zc_{{{\bf k}}\uparrow};c_{{{\bf k}}'\uparrow}^{\dag}
\rangle\rangle$ and
$\langle\langle S_-c_{{{\bf k}}\downarrow};c_{{{\bf k}}'\uparrow}^{\dag}
\rangle\rangle$ are similarly obtained as
\begin{align}
&(i\omega_n-v_zk_z+\mu)\langle\langle S_zc_{{{\bf k}}\uparrow};
c_{{{\bf k}}'\uparrow}^{\dag}\rangle\rangle_{i\omega_n}=
\langle S_z\rangle\delta_{{\bf k}{\bf k}'}\nonumber\\
&+ (v_xk_x-iv_yk_y)
\langle\langle S_zc_{{{\bf k}}\downarrow};c_{{{\bf k}}'\uparrow}^{\dag}
\rangle\rangle_{i\omega_n}\nonumber\\
&- \frac{J}{2N}\sum_{{\bf q}}\Big[ \langle\langle S_z^2c_{{\bf q}\uparrow};
c_{{\bf k}'\uparrow}^{\dag}\rangle\rangle_{i\omega_n}
+\frac{1}{2}\langle\langle S_{-}c_{{\bf q}\downarrow};
c_{{\bf k}'\uparrow}^{\dag}\rangle\rangle_{i\omega_n}\Big]\nonumber\\
&- \frac{J}{2N}\sum_{{\bf q}{\bf q}'}\Big[
\langle\langle S_{+}c_{{\bf k}\uparrow}c_{{\bf q}\downarrow}^{\dag}
c_{{\bf q}'\uparrow};c_{{\bf k}'\uparrow}^{\dag}\rangle\rangle_{i\omega_n}
\nonumber\\
& ~~ -\langle\langle S_{-}c_{{\bf k}\uparrow}c_{{\bf q}\uparrow}^{\dag}
c_{{\bf q}'\downarrow};c_{{\bf k}'\uparrow}^{\dag}\rangle\rangle_{i\omega_n}
\Big],
\label{szcc}
\end{align}
\begin{align}
&(i\omega_n+v_zk_z+\mu)\langle\langle S_{-}c_{{\bf k}\downarrow};
c_{{\bf k}'\uparrow}^{\dag}\rangle\rangle_{i\omega_n}\nonumber\\
&= (v_xk_x+iv_yk_y)
\langle\langle S_-c_{{{\bf k}}\uparrow};c_{{\bf k}'\uparrow}^{\dag}
\rangle\rangle_{i\omega_n}\nonumber\\
&- \frac{J}{2N}\sum_{{\bf q}}\Big[
\langle\langle S_{+}S_{-}c_{{\bf q}\uparrow};
c_{{\bf k}'\uparrow}^{\dag}\rangle\rangle_{i\omega_n}
+\frac{1}{2}\langle\langle S_{-}c_{{\bf q}\downarrow};
c_{{\bf k}'\uparrow}^{\dag}\rangle\rangle_{i\omega_n}\Big]\nonumber\\
&- \frac{J}{2N} \sum_{{\bf q}{\bf q}'}\Big[
\langle\langle S_{-}c_{{\bf k}\downarrow}c_{{\bf q}\uparrow}^{\dag}
c_{{\bf q}'\uparrow};
c_{{\bf k}'\uparrow}^{\dag}\rangle\rangle_{i\omega_n}\nonumber\\
&- 
\langle\langle S_{-}c_{{\bf k}\downarrow}c_{{\bf q}\downarrow}^{\dag}
c_{{\bf q}'\downarrow};
c_{{\bf k}'\uparrow}^{\dag}\rangle\rangle_{i\omega_n}\Big]\nonumber\\
&+ \frac{J}{N}\sum_{{\bf q}{\bf q}'}
\langle\langle S_{z}c_{{\bf k}\downarrow}c_{{\bf q}\downarrow}^{\dag}
c_{{\bf q}'\uparrow};
c_{{\bf k}'\uparrow}^{\dag}\rangle\rangle_{i\omega_n}.
\label{s-cc}
\end{align}

In this paper, we adopt the decoupling procedure for Green's functions including
several operators\cite{yan12,nag65,ham67,zit68}.  For example, 
\begin{align}
\langle\langle S_{+}c_{{\bf k}\uparrow}c_{{\bf q}\downarrow}^{\dag}
c_{{\bf q}'\uparrow};c_{{\bf k}'\uparrow}^{\dag}\rangle\rangle &\simeq
\langle S_{+}c_{{\bf q}\downarrow}^{\dag}c_{{\bf q}'\uparrow}\rangle
\langle\langle c_{{\bf k}\uparrow};c_{{\bf k}'\uparrow}^{\dag}
\rangle\rangle \nonumber\\
 &- \langle S_{+}c_{{\bf q}\downarrow}^{\dag}c_{{\bf k}\uparrow}\rangle
\langle\langle c_{{\bf q}'\uparrow};c_{{\bf k}'\uparrow}^{\dag}
\rangle\rangle .
\end{align}
We need further the Green's functions
$\langle\langle S_zc_{{{\bf k}}\downarrow};c_{{{\bf k}}'\uparrow}^{\dag}
\rangle\rangle$ and
$\langle\langle S_-c_{{{\bf k}}\uparrow};c_{{{\bf k}}'\uparrow}^{\dag}
\rangle\rangle$. 
We neglect the terms of the order of $J$ in the equations of motion for 
these Green's functions; this means that
we use the following approximation:
\begin{align}
\langle\langle S_zc_{{{\bf k}}\downarrow};c_{{{\bf k}}'\uparrow}^{\dag}
\rangle\rangle_{i\omega_n} &\simeq \frac{v_xk_x+iv_yk_y}{i\omega_n+v_zk_z+\mu}
\langle\langle S_zc_{{{\bf k}}\uparrow};c_{{{\bf k}}'\uparrow}^{\dag}
\rangle\rangle_{i\omega_n},\\
\langle\langle S_-c_{{{\bf k}}\uparrow};c_{{{\bf k}}'\uparrow}^{\dag}
\rangle\rangle_{i\omega_n} &\simeq \frac{v_xk_x-iv_yk_y}{i\omega_n-v_zk_z+\mu}
\langle\langle S_-c_{{{\bf k}}\downarrow};c_{{{\bf k}}'\uparrow}^{\dag}
\rangle\rangle_{i\omega_n}.
\end{align}
For $F_{{\bf k}{\bf k}'}=\langle\langle c_{{\bf k}\downarrow};
c_{{\bf k}'\uparrow}^{\dag}\rangle\rangle_{i\omega_n}$,
we neglect the terms of the order of $J$ in a similar way
to obtain
\begin{equation}
F_{{\bf k}{\bf k}'}(i\omega_n)\simeq \frac{(v_xk_x+iv_yk_y)}{i\omega_n+v_zk_z+\mu}
G_{{\bf k}{\bf k}'\uparrow}(i\omega_n)+O(J).
\end{equation}
We set $\langle S_z\rangle=0$.
From eqs.(\ref{szcc}) and (\ref{s-cc}), $\Gamma_{{\bf k}{\bf k}'\uparrow}$ 
is written as
\begin{align}
\Gamma_{{\bf k}{\bf k}'\uparrow}(i\omega_n)&= 
\frac{i\omega_n+\mu}{ (i\omega_n+\mu)^2-(v_x^2k_x^2+v_y^2k_y^2+v_z^2k_z^2)}
\nonumber\\
&\times \Big[ -\left(\frac{3}{4}-m_{{\bf k}}\right)\frac{J}{2N}
G_{{\bf k}'\uparrow}^0(i\omega_n)\nonumber\\
&~~ + \left( \frac{3}{4}-m_{{\bf k}}\right)\frac{J}{2N}\sum_{{\bf q}}
G_{{\bf q}}^0(i\omega_n)\frac{J}{2N}\sum_{{\bf p}}\Gamma_{{\bf p}{\bf k}'\uparrow}
(i\omega_n)\nonumber\\
&~~ -\Big( n_{{\bf k}\uparrow}+n_{{\bf k}\downarrow}-1\Big)\frac{J}{2N}
\sum_{{\bf p}}
\Gamma_{{\bf p}{\bf k}'\uparrow}(i\omega_n) \Big]+\cdots,
\end{align}
where $\cdots$ indicates the terms, being proportional to $k_z$, which give
small contributions to $G_{{\bf k}{\bf k}'\uparrow}$ because they would vanish
when we take the summation with respect to ${\bf k}$.
Here we set 
\begin{align}
m_{{\bf k}}&=3\sum_{{\bf q}} \langle c_{{\bf q}\uparrow}^{\dag}c_{{\bf k}\downarrow}
S_{-}\rangle,\\
n_{{\bf k}\sigma}&=\sum_{{\bf q}}\langle c_{{\bf q}\sigma}^{\dag}c_{{\bf k}\sigma}\rangle,
\end{align}
and define 
\begin{align}
G_{{\bf k}}^0(i\omega_n)&=\frac{i\omega_n+\mu}{ (i\omega_n+\mu)^2-(v_x^2k_x^2
+v_y^2k_y^2+v_z^2 k_z^2)},\\
G_{{\bf k}\sigma}^0(i\omega_n)&=\frac{i\omega_n+\mu+\sigma v_zk_z}{ 
(i\omega_n+\mu)^2-(v_x^2k_x^2
+v_y^2k_y^2+v_z^2 k_z^2)}.\\
\end{align}
We also define the following functions:
\begin{align}
F(i\omega_n) &= \frac{1}{N}\sum_{{\bf k}}G_{{\bf k}}^0(i\omega_n),\\
\Gamma(i\omega_n) &= \frac{1}{N}\sum_{{\bf k}} \left(m_{{\bf k}}-\frac{3}{4}\right)
G_{\bf k}^0(i\omega_n),\\
G(i\omega_n) &= \frac{1}{2N}\sum_{{\bf k}} \left( n_{{\bf k}\uparrow}
+n_{{\bf k}\downarrow}-1\right) 
G_{{\bf k}}^0(i\omega_n).
\end{align}
Then we obtain
\begin{equation}
\sum_{{\bf k}}\Gamma_{{\bf k}{\bf k}'\uparrow}=\frac{J}{2}
\frac{\Gamma(i\omega_n)G^0_{{\bf k}'\uparrow}(i\omega_n)}
{1+JG(i\omega_n)+\left(\frac{J}{2}\right)^2\Gamma(i\omega_n)F(i\omega_n)}.
\end{equation}
Now we can obtain a closed solution
for a set of Green's functions.
The Green's function $G_{{{\bf k}}{{\bf k}}'\uparrow}(i\omega_n)$ reads
\begin{align}
G_{{{\bf k}}{{\bf k}}'\uparrow}(i\omega_n) &=  
\delta_{{\bf k}{\bf k}'}G_{{\bf k}\uparrow}^0(i\omega_n)- 
\frac{J}{2N}G_{{\bf k}\uparrow}^0(i\omega_n)\frac{J}{2}
\Gamma(i\omega_n)\nonumber\\
&~ \times G_{{\bf k}'\uparrow}^0(i\omega_n)
\frac{1}{1+JG(i\omega_n)+\left( \frac{J}{2}\right)^2
\Gamma(i\omega_n)F(i\omega_n)}.
\end{align}
The Green's function $G_{{\bf k}{\bf k}'\downarrow}(i\omega_n)$ is
similarly obtained as
\begin{align}
G_{{{\bf k}}{{\bf k}}'\downarrow}(i\omega_n) &=  
\delta_{{\bf k}{\bf k}'}G_{{\bf k}\downarrow}^0(i\omega_n)- 
\frac{J}{2N}G_{{\bf k}\downarrow}^0(i\omega_n)\frac{J}{2}
\Gamma(i\omega_n)\nonumber\\
&~ \times G_{{\bf k}'\downarrow}^0(i\omega_n)
 \frac{1}{1+JG(i\omega_n)+\left( \frac{J}{2}\right)^2
\Gamma(i\omega_n)F(i\omega_n)}.
\end{align}

\section{Kondo Temperature}

From the Green's function $G_{{{\bf k}}{{\bf k}}'\uparrow}(i\omega_n)$, the Kondo 
temperature $T_{\rm K}$ is
determined from a zero of the denominator in this formula.  
We perform the analytic continuation $i\omega_n\rightarrow \omega$ and
consider
\begin{equation}
1+JG(\omega)=0
\label{tkeq}
\end{equation}
in the limit $\omega\rightarrow 0$ by neglecting higher-order term being 
proportional to $(J/2)^2$.
We neglect the term of the order of $J$ in 
$n_{{\bf k}\sigma}=\langle c_{{\bf k}\sigma}^{\dag}
c_{{\bf k}\sigma}\rangle$.
The equation in eq.(\ref{tkeq}) is written as
\begin{equation}
1=\frac{1}{4}J\frac{1}{N}\sum_{{\bf k}}G_{{\bf k}}^0(\omega)\Big[
\tanh\left(\frac{\epsilon_{{\bf k}}-\mu}{2k_{\rm B}T_{\rm K}}\right)
+\tanh\left(\frac{-\epsilon_{{\bf k}}-\mu}{2k_{\rm B}T_{\rm K}}\right)\Big],
\end{equation}
where $\epsilon_{{\bf k}}=\sqrt{v_x^2k_x^2+v_y^2k_y^2+v_z^2k_z^2}$.
Let us adopt for simplicity that $v_x=v_y=v_z=v$ and the dispersion is
$\xi_k=\pm v\sqrt{k_x^2+k_y^2+k_z^2}-\mu$.
We can consider, in general, the $d$-dimensional case; by setting
$v_z=0$ for two dimensions and $v_y=v_z=0$ for one dimension.
Then the equation for $T_{\rm K}$ is
\begin{align}
1&= \frac{1}{8}J\frac{\Omega_d}{(2\pi)^d}\int_0^{D} dk k^{d-1}
\left( \frac{1}{\omega+\mu-vk}+\frac{1}{\omega+\mu+vk} \right)\nonumber\\
&\times T_{\rm K}\sum_{n=-\infty}^{\infty}\Big[
\frac{1}{vk-\mu-i\pi(2n+1)T_{\rm K}}\nonumber\\
&~~~~~~~~~~ -\frac{1}{vk+\mu+i\pi(2n+1)T_{\rm K}} \Big],
\label{tkeq2}
\end{align}
where $\Omega_d$ is the solid angle in $d$-dimensional space, namely, the
area of the $(d-1)$-sphere $S^{d-1}$. $D$ is an cutoff and we set $k_{\rm B}=1$
for simplicity.
When $d$ is odd, the integrand is an even function of $k$.
In this case the integral with respect to $k$ is reduced to an
integral in the complex plane.
For $d=3$, the equation in eq.(\ref{tkeq2}) is written as
\begin{align}
1&= \frac{1}{32\pi^2}J \Big[ \sum_{n\geq 0}\frac{1}{v}
\left( \frac{\mu+i\pi(2n+1)T_{\rm K}}{v}\right)^2\nonumber\\
&\times 2\pi iT_{\rm K}\left( \frac{1}{\omega-i\pi(2n+1)T_{\rm K}}
+\frac{1}{\omega+2\mu+i\pi(2n+1)T_{\rm K}}\right) \nonumber \\
& -\sum_{n<0}\frac{1}{v}\left( \frac{\mu+i\pi(2n+1)T_{\rm K}}{v}\right)^2
\nonumber\\
&\times 2\pi iT_{\rm K}\left( \frac{1}{\omega-i\pi(2n+1)T_{\rm K}}
+\frac{1}{\omega+2\mu+i\pi(2n+1)T_{\rm K}}\right)\Big],
\end{align}
where the summation has the upper limit $n_0\equiv D/(2\pi T_{\rm K})$.
By using the formula of the digamma function,
\begin{equation}
\sum_{n=0}^{n_0}\frac{1}{n+\frac{1}{2}+x}=
\psi\left(\frac{1}{2}+x+n_0\right)-\psi\left(\frac{1}{2}+x\right),
\end{equation}
we obtain
\begin{align}
1&= \frac{1}{32\pi^2}|J|\frac{\mu^2}{v^3}\Bigg[
\psi\left( \frac{1}{2}-\frac{\omega}{2\pi iT_{\rm K}}+n_0 \right)
-\psi\left( \frac{1}{2}-\frac{\omega}{2\pi iT_{\rm K}}\right) \nonumber\\
&~~ -\psi\left( \frac{1}{2}+\frac{\omega+2\mu}{2\pi iT_{\rm K}}+n_0 \right)
+\psi\left( \frac{1}{2}+\frac{\omega+2\mu}{2\pi iT_{\rm K}}\right) \nonumber\\
&~~ +\psi\left( \frac{1}{2}+\frac{\omega}{2\pi iT_{\rm K}}+n_0 \right)
-\psi\left( \frac{1}{2}+\frac{\omega}{2\pi iT_{\rm K}}\right) \nonumber\\
&~~ -\psi\left( \frac{1}{2}-\frac{\omega+2\mu}{2\pi iT_{\rm K}}+n_0 \right)
+\psi\left( \frac{1}{2}-\frac{\omega+2\mu}{2\pi iT_{\rm K}}\right) \Bigg].
\end{align}

We assume that the cutoff $D$ is much larger than the temperature: $D\gg T$. 
We employ the asymptotic form of $\psi(z)\sim \log (z)$ for large $z$
Then the $T_{\rm K}$ is the solution of the equation
\begin{equation}
1=\frac{1}{16}\rho_{\rm d}(\mu)|J|\Bigg[K(t)
+\log\left(\frac{D}{\sqrt{D^2+4\mu^2}}\right)\Bigg],
\label{tkeq3}
\end{equation}
where we defined $t\equiv T/|\mu|$ and
\begin{equation}
K(t)={\rm Re}\psi\left(\frac{1}{2}+i\frac{1}{\pi t}\right)
-\psi\left(\frac{1}{2}\right).
\end{equation}
We introduced the density of states $\rho_{\rm d}$ as
\begin{equation}
\rho_{\rm d}(\mu)= \frac{\Omega_{\rm d}}{(2\pi)^d}
\left(\frac{|\mu|}{v}\right)^{d-1}\frac{1}{v}.
\end{equation}
Here, $\Omega_d$ is the solid angle in $d$-dimensional space, that is, 
the area of the $(d-1)$-sphere $S^{d-1}$.
As $t$ approaches 0, $K(t)$ behaves as $K(t)\sim \log(1/\pi t)$.
For large $t$, $K(t)\sim 7\zeta(3)/(\pi^2 t^2)$ where $\zeta(3)$ is the
Riemann zeta function at argument 3.
The equation eq.(\ref{tkeq3}) always has a solution for $|J|>0$. 
When $\rho_{\rm d}|J|$ is small, $\rho_{\rm d}|J|\ll 1$, we have a solution
in the logarithmic region of $K(t)$.  Then we obtain the Kondo
temperature,
\begin{equation}
k_{\rm B} T_{\rm K} = \frac{2e^{\gamma}D}{\pi}\frac{1}{\sqrt{1+\frac{D^2}{4\mu^2}}}
\exp\left( -\frac{8}{\rho_{\rm d}(\mu)|J|}\right),
\label{tkformula}
\end{equation}
where $\gamma$ is Euler's constant and $k_{\rm B}$ is included in the formula
of $T_{\rm K}$.
The result shows that the Kondo effect indeed occurs in a Dirac system.

When $D\gg |\mu|$, $T_{\rm K}$ is
\begin{equation}
k_{\rm B} T_{\rm K} = \frac{4e^{\gamma}|\mu|}{\pi}
\exp\left( -\frac{8}{\rho_{\rm d}(\mu)|J|}\right).
\end{equation}
In the limit $|\mu|\rightarrow 0$, the equation $1+JG(0)=0$ has no solution.
This indicates that when $|\mu|$ is small, $T_{\rm K}$ is reduced and vanishes
for $|\mu|\rightarrow 0$.
Hence, when the Fermi surface is point like,
the Kondo effect never appears; this is because the scattering
from the localized spin becomes weak for the point-like Fermi surface.
This is consistent with calculation obtained by using the Abrikosov-fermion
mean field theory for a topological insulator\cite{ori13} and that by the 
functional-integral saddle-point
theory\cite{pri14}.

We also examine the case where $\rho_{\rm d}|J|$ is large being of order 1 although
our method is likely, however, not reliable in this region.
In this region, we have a solution for $K(t)\sim 7\zeta(3)/(\pi^2t^2)+O(1/t^4)$ and
$T_{\rm K}$ is an algebraic function $\mu$ such as $\mu^{\alpha}$ with a constant
$\alpha$.  If we use $K(t)\simeq 7\zeta(3)/(\pi^2 t^2)$, we obtain
\begin{align}
k_{\rm B} T_{\rm K} &= \frac{1}{\pi^2}\sqrt{\frac{7\zeta(3)}{16}}\mu^2\frac{1}{v}
\sqrt{\frac{|J|}{v}}\nonumber\\
&= \frac{1}{\pi}\sqrt{\frac{7\zeta(3)}{8}}|\mu|\sqrt{\rho_{\rm d}(\mu)|J|}.
\label{tkjl}
\end{align}
$T_{\rm K}$ is proportional to $|\mu|$ times an algebraic function of $\rho_{\rm d}|J|$.

In one dimension ($d=1$), $T_K$ is obtained in a similar way.  
For small $\rho_{\rm d}|J|$,  we have
\begin{equation}
k_{\rm B} T_{\rm K} = \frac{2e^{\gamma}D}{\pi}\frac{1}{\sqrt{1+\frac{D^2}{4\mu^2}}}
\exp\left( -8\pi v \frac{1}{|J|} \right).
\end{equation}
For small $\mu$, $T_{\rm K}$ is 
\begin{equation}
k_{\rm B} T_{\rm K} = \frac{4e^{\gamma}|\mu|}{\pi}
\exp\left( -8\pi v \frac{1}{|J|} \right).
\end{equation}
Let us turn to the two-dimensional case.
The integrand is an odd function of $k$ for $d=2$, and thus the integral
is not straightforwardly reduced to a complex integral.
The results for $d=3$ and 1, however, show that $k^{d-1}$ in the integrand is
approximately replaced by $(|\mu|/v)^{d-1}$ because the zero of
denominators in the integrand gives important contributions.
This results in the formula of $T_{\rm K}$ for $d=2$. 

As a result, the formula of $T_{\rm K}$ in $d$ dimensions reads as in
eq.(\ref{tkformula}) for $d$=1, 2, $\cdots$.
The solution for $\rho_{\rm d}|J|$ being of order 1 in the region
$K(t)\sim 7\zeta(3)/(\pi^2 t^2)$ is given by eq.(\ref{tkjl}).

We derived $T_{\rm K}$ by using the Green's function theory.
$T_{\rm K}$ is proportional to $|\mu|$ with the exponential factor
for small $\rho_{\rm d}|J|$ and $|\mu|\ll D$.
The formula of $T_{\rm K}$ shows that $T_{\rm K}$ decreases as the 
dimension $d$ is increased.
We summarize the results for $T_{\rm K}$ in Table 1 (the first column).

There has been a calculation based on the numerical
renormalization group method\cite{wil75} for a pseudogap $U=\infty$
Anderson model with the density of states $\rho(\omega)\propto |\omega+\mu|^r$
for $r=1$\cite{voj10}.
Their results indicate a particle-hole asymmetry showing
$T_{\rm K}\propto \mu^x$ with $x\simeq 2.6$ for $\mu>0$ and 
$T_{\rm K}\propto |\mu|$ for $\mu<0$.
This kind of asymmetry cannot be visible in our method.

\begin{table}
\caption{Characteristic behaviors of
Kondo temperature $T_{\rm K}$, the resistivity $R$ 
and the specific heat $\Delta C$
where $\bar{D}=D/\sqrt{1+D^2/(2\mu)^2}$.
We set $K(t)={\rm Re}\psi(1/2+i\mu/(\pi T))-\psi(1/2)$ and
$I(t)={\rm Im}\psi'(1/2+i\mu/(\pi T))$, where $\psi(z)$ is the
digamma function.  $\alpha$ is a constant.  
We assume that $T\ll D$.
}
\newlength{\myheight}
\setlength{\myheight}{1.0cm}
\begin{center}
\begin{tabular}{|c|c|c|}
\hline
Quantities & Conditions   & \\
\hline
\parbox[c][\myheight][c]{0cm}{}
   & $\rho_{\rm d}|J|\ll 1$    &  
   $T_{\rm K}\simeq \bar{D}\exp\left( -\frac{8}{\rho_d(\mu)|J|}\right)$ \\
\parbox[c][\myheight][c]{0cm}{}
$T_{\rm K}$   &   $~~~~(|\mu|\ll D)$  &
    $T_{\rm K}\simeq |\mu|\exp\left( -\frac{8}{\rho_d(\mu)|J|}\right)$ \\
\cline{2-3}
\parbox[c][\myheight][c]{0cm}{}
   &  $\rho_{\rm d}|J|\sim O(1)$ & $T_{\rm K}\simeq |\mu|(\sqrt{\rho_{\rm d}|J|}+\cdots)$  \\
\hline
\parbox[c][\myheight][c]{0cm}{}
    & $T_{\rm K}<T\ll |\mu|$ &   
$R\simeq \frac{c\pi}{e^2v^2\rho_d}|J|
\left(\log\left(\frac{T}{T_{\rm K}}\right)\right)^{-1}$  \\
\parbox[c][\myheight][c]{0cm}{}
$R$  & & $~\simeq \frac{c\pi}{e^2v^2}\frac{1}{8}|J|^2\left( 1+\frac{1}{8}
\rho_{\rm d}|J|\log\frac{\bar{D}}{T}\right)$ \\
\cline{2-3}
\parbox[c][\myheight][c]{0cm}{}
     &  $T\sim |\mu|$ &  
 $R \simeq \frac{c\pi}{e^2v^2}|J|^2\left( 1+\frac{1}{8}\rho_{\rm d}|J|K(t)\right)$  \\
\cline{2-3}
\parbox[c][\myheight][c]{0cm}{}
  &  $|\mu|\ll T$  &   
 $R \simeq \frac{c\pi}{e^2v^2}|J|^2$  \\
\hline
\parbox[c][\myheight][c]{0cm}{}
    & $T_{\rm K}<T\ll |\mu|$ &   
$\Delta C\propto \left(\frac{1}{8}\rho_{\rm d}|J|\right)^{3/2}T
\left(\log\frac{T}{T_{\rm K}}\right)^{-3/2}$  \\
\parbox[c][\myheight][c]{0cm}{}
$\Delta C$  &  & $~\propto \left(\frac{1}{8}\rho_{\rm d}|J|\right)^3T
\left( 1+\frac{3}{16}\rho_{\rm d}|J|\log\frac{\bar{D}}{T} \right)$ \\
\cline{2-3}
\parbox[c][\myheight][c]{0cm}{}
  &  $ T\sim |\mu|$ &  
 $\Delta C \propto T\frac{\partial^2}{\partial T^2}\left(C_1T K(t)+C_2 TI(t) \right)$  \\
\cline{2-3}
\parbox[c][\myheight][c]{0cm}{}
  &  $|\mu|\ll T$  &   
 $\Delta C \simeq  O(\mu/T)$  \\
\hline
\end{tabular}
\end{center}
\end{table}

\section{Electrical Resistivity}

We consider the conductivity given by the formula:
\begin{equation}
\sigma = -\frac{2e^2}{3}\int \tau_k v_k^2\frac{\partial f}{\partial\xi_k}\rho_d
d\xi_k.
\end{equation}
The life time $\tau_k$ is given as
\begin{equation}
\frac{1}{\tau_k}= cN{\rm Im}G_{{\bf k}{\bf k}\uparrow}(\omega+i\delta)^{-1},
\end{equation}
where $c$ is the concentration of magnetic impurities.
We adopt that $T>T_K$ so that $m_k=0$.  Then we have 
$\Gamma(\omega)=-(3/4)F(\omega)$
and the Green function is
\begin{align}
G_{{\bf k}{\bf k}\uparrow}(\omega)&= G_{{\bf k}\uparrow}^0(\omega)\Bigg[
1-\left( \frac{J}{2}\right)^2\frac{3}{4N}F(\omega)G_{{\bf k}\uparrow}^0(\omega)
\nonumber\\
&~ \times \frac{1}{1+JG(\omega)-\left(\frac{J}{2}\right)^2\frac{3}{4}F(\omega)^2}
\Bigg].
\end{align}
This is approximately given as
\begin{equation}
G_{{\bf k}{\bf k}\uparrow}(\omega)^{-1}\simeq G_{{\bf k}\uparrow}^0(\omega)^{-1}
-\frac{3J^2}{16N}\frac{F(\omega)}{1+JG(\omega)}+O(J^4).
\end{equation}
${\rm Im}F(\omega+i\delta)$ is the density of states:
\begin{equation}
{\rm Im}F(\omega+i\delta)=-\frac{1}{2}\pi \rho_{\rm d}(\omega+\mu),
\end{equation}
where we assume that $\mu>0$.  Then the life time $\tau_k$ reads
\begin{equation}
\tau_k(\omega)\simeq \frac{32}{3c\pi J^2\rho_{\rm d}(\omega+\mu)}(1+JG(\omega)).
\end{equation}
This results in the conductivity:
\begin{align}
\sigma &\simeq 
 \frac{2e^2}{3}v^2\frac{32}{3c\pi J^2}\Bigl(1-|J|
G(0)\Bigr)\nonumber\\
&\simeq \frac{2e^2}{3}v^2\frac{32}{3c\pi J^2}\Biggl( 1+\frac{1}{16}
\rho_{\rm d}(\mu)|J|
\nonumber\\
&~ \times \Bigg[ -2\psi\left( \frac{1}{2}+\frac{D}{2\pi T}
\right)+2\psi\left( \frac{1}{2} \right)\nonumber\\
&~ + \psi\left( \frac{1}{2}+\frac{\mu}{\pi iT}+\frac{D}{2\pi T}
\right)-\psi\left( \frac{1}{2}+\frac{\mu}{\pi iT}\right)\nonumber\\
&~ + \psi\left( \frac{1}{2}-\frac{\mu}{\pi iT}+\frac{D}{2\pi T}
\right)-\psi\left( \frac{1}{2}-\frac{\mu}{\pi iT}\right) \Bigg] \Biggr).
\nonumber\\
\label{conduct}
\end{align}

We assume $T\ll D$.
At low temperatures, when $T\ll |\mu|$, we obtain
\begin{equation}
\sigma\simeq \frac{8}{9c\pi}e^2v^2\frac{1}{|J|}\rho_{\rm d}(\mu)
\log\left(\frac{T}{T_{\rm K}}\right).
\end{equation}
The resistivity $R=1/\sigma$ is
\begin{align}
R&= \frac{9}{64}c\pi\frac{1}{e^2v^2}|J|^2 \Bigg[
1-\frac{\rho_{\rm d}(\mu)|J|}{8}\log\left( \frac{2e^{\gamma}\bar{D}}{\pi T}
 \right) \Bigg]^{-1}\nonumber\\
&\simeq \frac{9}{64}c\pi\frac{1}{e^2v^2}|J|^2\Bigg[
1+\frac{\rho_{\rm d}(\mu)|J|}{8}\log\left( \frac{2e^{\gamma}\bar{D}}{\pi T} \right)
\Bigg],
\end{align}
with $\bar{D}=D/\sqrt{1+D^2/(2\mu)^2}$.
There is a logarithmic $T$-dependence in the resistivity, which 
characterizes the Kondo effect.
This is consistent with the result in Ref.\cite{wan13} up to log-$T$ term.
We must note that the coefficient of the logarithmic term depends on
the chemical potential $\mu$.

When $T\sim O(|\mu|)$, $R$ reads
\begin{align}
R&\simeq \frac{9}{64}c\pi\frac{1}{e^2v^2}|J|^2\Biggl( 1+\frac{1}{8}
\rho_{\rm d}(\mu)|J|\Bigg[ K(t)\nonumber\\
&+ \log\left( \frac{D}{\sqrt{D^2+4\mu^2}}\right)\Bigg]\Biggr).
\end{align}
The log-$T$ dependence is reduced in this region.
In the high-temperature region where $|\mu| \ll T$ holds, the log($T$) term 
will be absent.
In the high-temperature region for $|\mu|\ll T$, the resistivity is
\begin{align}
R&\simeq \frac{9}{64}c\pi\frac{1}{e^2v^2}|J|^2\Biggl( 1+\frac{1}{8}
\rho_{\rm d}(\mu)|J|\nonumber\\
&~ \times \Bigg[ 7\zeta(3)\frac{1}{\pi^2}\left(\frac{\mu}{T}\right)^2
-31\zeta(5)\frac{1}{\pi^4}\left(\frac{\mu}{T}\right)^4+\cdots \Bigg]
\Biggr).
\end{align}
This has no log-$T$ dependence.
As a result log-$T$ appears in the region $T_{\rm K}< T\ll |\mu|$.
We summarize the results for $R$ in the second column of Table 1.
The log-$T$ dependence is reduced as the chemical potential $|\mu|$ 
approaches the Dirac point, that is, log-$T$ will disappear as $\mu\rightarrow 0$.

\section{Specific Heat}

Let us examine the specific heat in this section.
The specific heat also shows a singularity as in other physical quantities.
We calculate the additional entropy coming from the s-d interaction
with magnetic impurities.
The expectation value of the interaction part $H_{sd}$ is given by
\begin{align}
V &\equiv \langle H_{sd}\rangle \nonumber\\
& = \frac{J}{2N\beta}\sum_{{\bf k}{\bf k}'n\sigma}
\Gamma_{{\bf k}{\bf k}'\sigma}(i\omega_n)\nonumber\\
&= \frac{J}{N\beta}\sum_{{\bf k}n\sigma}G_{{\bf k}\sigma}^0(i\omega_n)
t(i\omega_n)
= \frac{2J}{N\beta}\sum_{{\bf k}n}G_{{\bf k}}^0(i\omega_n)t(i\omega_n),
\end{align}
where $t(z)$ is defined as
\begin{equation}
t(z)= -\frac{J}{4}\frac{\Gamma(z)}{1+JG(z)+(J/2)^2\Gamma(z)F(z)}.
\end{equation}
The $\omega_n$-summation is performed to give
\begin{align}
V&= 2J\frac{1}{N}\sum_{{\bf k}}\frac{1}{2\pi i}\int_{-\infty}^{\infty}
d\omega f(\omega)\Bigg[ \nonumber\\ 
& {\rm Re}\left( \frac{1}{\omega+\mu+vk}
+\frac{1}{\omega+\mu-vk} \right)
{\rm Im}t(\omega-i\delta)\nonumber\\
&+ i\pi\Bigl( \delta(\omega+\mu+vk)+\delta(\omega+\mu-vk)\Bigr)
{\rm Re}t(\omega-i\delta) \Bigg].
\end{align}
Because the $\omega$-dependence in $t(\omega)$ is important, we
neglect the $\omega$-dependence in $1/(\omega+\mu\pm vk)$ and 
$\delta(\omega+\mu\pm vk)$.
We adopt $m_k=0$ ($T>T_{\rm K}$), so that 
$\Gamma(\omega-i\delta)=-(3/4)F(\omega-i\delta)$.
$F(\omega-i\delta)$ is approximated as
$F(\omega-i\delta)\simeq F(-i\delta)\simeq 
\rho_{\rm d}(\mu)F_{\rm d}+i\pi\rho_{\rm d}(\mu)/2$, 
where $F_{\rm d}=-{\rm sign}(\mu)v\Lambda/|\mu|$ for $d=3$ and
$F_{\rm d}= -{\rm sign}(\mu)(1/2)\ln|(v^2\Lambda^2-\mu^2)/\mu^2|$ for $d=2$
with a cutoff $\Lambda$ in $k$-integration.
The real part of $F(\omega-i\delta)$ is proportional to $\mu/|\mu|$.
Then the interaction energy is
\begin{equation}
V = \frac{3}{8}F_{\rm d}\Bigg(\rho_{\rm d}(\mu)J\Bigg)^2
\int_{-\infty}^{\infty}d\omega f(\omega)\frac{1}{1+JG(\omega)}.
\end{equation}
Using the relation between the free energy and $V$ given as
\begin{equation}
V = J\frac{\partial F}{\partial J},
\end{equation}
the additional entropy is obtained as
\begin{equation}
\Delta S = -\frac{\partial}{\partial T}(F-F_0)
=-\int_0^J \frac{dJ'}{J'}\frac{\partial}{\partial T}V(J',T).
\end{equation}

We employ the expansion formula for the Fermi distribution function
$f(\omega)$:
\begin{equation}
\int_D^D d\omega f(\omega)h(\omega)= \int_D^0 d\omega h(\omega)
+\frac{\pi^2}{6}(k_BT)^2h'(0),
\end{equation}
for a differentiable function $h(\omega)$.  Then we have
\begin{align}
V &\simeq \frac{3}{8}F_{\rm d}\Big(\rho_{\rm d}(\mu)J\Big)^2
\Bigg[ \int_{-D}^0 d\omega\frac{1}{1+JG(\omega)}\nonumber\\
& ~~ +\frac{\pi^2}{6}(k_B T)^2\frac{\partial}{\partial\omega}
\frac{1}{1+JG(\omega)}\Big|_{\omega=0} \Bigg].
\label{vexpan}
\end{align}
We first examine the low-temperature region characterized by
$T\ll |\mu|$.
When $|\mu|/T$ is large, the second term of $V$ is written as
\begin{equation}
V_2= -\frac{\pi^2}{2}F_{\rm d}\rho_{\rm d}(\mu)|J|(k_{\rm B} T)^2
\frac{1}{\Big(\ln(T_{\rm K}/T)-g(\omega)\Big)^2}g'(\omega)\Bigg|_{\omega=0},
\end{equation}
where $g(\omega)$ is defined as
\begin{equation}
g(\omega)= \ln\left(\frac{T_{\rm K}}{T}\right)+\frac{8}{\rho_{\rm d}(\mu)|J|}
\Big( 1+JG(\omega)\Big).
\end{equation}
We have $g(0)=0$ and $g'(0)$ is evaluated as
\begin{align}
g'(0)&\simeq -\frac{1}{2}\Bigg[ \frac{1}{2\pi iT}\psi'\left( \frac{1}{2}
+\frac{\mu}{\pi iT}\right)-\frac{1}{2\pi iT}\psi'\left( \frac{1}{2}
-\frac{\mu}{\pi iT} \right)\nonumber\\
&+ \frac{1}{2\pi iT}\psi'\left( \frac{1}{2}-\frac{\mu}{\pi iT}+\frac{D}{2\pi T}
\right)\nonumber\\
&-\frac{1}{2\pi iT}\psi'\left( \frac{1}{2}+\frac{\mu}{\pi iT}
+\frac{D}{2\pi T}\right) \Bigg]\nonumber\\
&\simeq -\frac{1}{2\mu}\frac{D^2}{D^2+4\mu^2}.
\end{align}
This is in contrast to the case of conventional Kondo effect where $g'(0)$ is
proportional to the inverse of temperature $1/T$.
Then we obtain
\begin{equation}
V_2 = \frac{\pi^2}{4}A_{\rm D}\rho_d(\mu)|J|\frac{F_{\rm d}}{\mu}(k_{\rm B} T)^2
\frac{1}{\Big( \ln(T_{\rm K}/T)\Big)^2},
\end{equation}
with $A_{\rm D}$ is a constant given by $D^2/(D^2+4\mu^2)$.
The first term of $V$ has a similar singularity\cite{yan12}. 
We expand this term in terms of the inverse of
$\ln(T_{\rm K}/T)$ to obtain
\begin{equation}
V_1= -3F_{\rm d}\rho_{\rm d}(\mu)|J|
\frac{1}{\left(\ln(T_{\rm K}/T)\right)^2}\int_{-D}^0 d\omega g(\omega)+\cdots.
\end{equation}
$g(\omega)$ is expanded by means of $\omega/T$ by restricting the integral
region to $(-k_{\rm B}T,0)$ to give a term $(k_{\rm B}T)^2/(\ln(T_{\rm K}/T))^2$.
This contribution is included in the coefficient $A_{\rm D}$ such as
$A_{\rm D}=(1-3/\pi^2)D^2/(D^2+4\mu^2)$ and we have
\begin{equation}
V = \frac{\pi^2}{4}A_{\rm D}\rho_{\rm d}(\mu)|J|\frac{F_{\rm d}}{\mu}(k_{\rm B} T)^2
\frac{1}{\Big( \ln(T_{\rm K}/T)\Big)^2}.
\end{equation}

Because of the relation 
$\ln T_{\rm K}=\ln(2e^{\gamma}\bar{D}/{\pi})+8/(\rho_{\rm d} J)$,
the following holds:
$d\ln T_{\rm K} = -8d(\rho_{\rm d} J)/(\rho_{\rm d} J)^2$.
Using this, the contribution to the free energy is evaluated as
\begin{align}
\Delta F &= \int_0^{\rho_{\rm d} J}\frac{d(\rho_{\rm d} J)}{\rho_{\rm d} J}
V(\rho_{\rm d} J)\nonumber\\
&= -2\pi^2 A_{\rm D}\frac{F_{\rm d}}{\mu}(k_{\rm B} T)^2\frac{1}{(\ln(\bar{D}/T))^2}
\Bigg[ \frac{\rho_{\rm d} J}{8}\nonumber\\
&+ \frac{1}{\ln(T_{\rm K}/T)}-\frac{2}{\ln(\bar{D}/T)}\ln\Bigg|
\frac{\rho_{\rm d} J}{8}\ln\left( \frac{T_{\rm K}}{T}\right)\Bigg| \Bigg],
\end{align}
where we neglected the factor $2e^{\gamma}/\pi$ for simplicity.
The entropy $\Delta S=-\partial\Delta F/\partial T$ is found to be
\begin{align}
\Delta S &= 4\pi^2 A_{\rm D} k_{\rm B}F_{\rm d}\frac{k_{\rm B} T}{\mu}\Bigg[ 
\frac{1}{3}\left(\frac{\rho_{\rm d} J}{8}\right)^3
-\frac{1}{2}\left(\frac{\rho_{\rm d} J}{8}\right)^4\ln\frac{\bar{D}}{T}
\nonumber\\
& -\frac{3}{5}\left(\frac{\rho_{\rm d} J}{8}\right)^5\ln\frac{\bar{D}}{T}
 +\frac{3}{5}\left(\frac{\rho_{\rm d} J}{8}\right)^5
\left(\ln\frac{\bar{D}}{T}\right)^2+\dots
\Bigg].
\end{align}
The entropy has a singularity such as $T\ln T$ in the fourth order of $\rho_d J$, 
which is rather weak
singularity compared to the conventional Kondo effect with $\ln T$ term.
This results in the specific heat $\Delta C=T\partial\Delta S/\partial T$:
\begin{align}
\frac{\Delta C}{k_{\rm B}}&\simeq \frac{4\pi^2}{3}A_{\rm D}F_{\rm d}
\frac{k_{\rm B} T}{\mu}
\left(\frac{\rho_{\rm d} J}{8}\right)^3\Bigg[ 1-\frac{3}{2}
\left(\frac{\rho_{\rm d} J}{8}\right)
\ln\frac{\bar{D}}{T}+\cdots \Bigg]\nonumber\\
&\simeq \frac{4\pi^2}{3}A_{\rm D}|F_{\rm d}| \frac{k_{\rm B} T}{|\mu|}
\left(\frac{\rho_{\rm d} |J|}{8}\right)^{3/2}
\frac{1}{\Big(\ln(T/T_{\rm K})\Big)^{3/2}}.
\end{align}
Hence, the specific heat in a Dirac system has a singularity as a function of
the temperature near $T_{\rm K}$.
This should be compared with the specific heat anomaly in the original
Kondo effect given as $\Delta C/k_{\rm B} \simeq 1/\Big( \ln(T_{\rm K}/T)\Big)^4$,
which is shown by following the above method for the original s-d model.
The singularity in a Dirac system is much weaker than that of the original 
Kondo effect.
The log-$T$ appears in the range $T_{\rm K} < T\ll |\mu|/k_{\rm B}$.

Let us then examine the specific heat in the intermediate region $T\sim |\mu|$. 
The expectation value of the interaction term $V$ in eq.(\ref{vexpan}) is evaluated by
expanding $JG(\omega)$ in $\omega$ and the integral is restricted to the 
interval $(-T,0)$.
$JG(\omega)$ is written as
\begin{equation}
JG(\omega)=JG(0)+\frac{1}{16}\rho_{\rm d}|J|\frac{\omega}{T}
{\rm Im}\psi'\left(\frac{1}{2}+i\frac{\mu}{\pi T}\right),
\end{equation}
and the main contributions to $V$ are
\begin{equation}
V\simeq F_{\rm d}(\rho_{\rm d}|J|)^3 \Bigg[ C_1 TK(t) + C_2 T
{\rm Im}\psi'\left(\frac{1}{2}+i\frac{\mu}{\pi T}\right) \Bigg],
\end{equation}
where $C_1$ and $C_2$ are constants.
This results in the specific heat given as
\begin{equation}
\frac{\Delta C}{k_{\rm B}}\simeq -\frac{1}{3}F_{\rm d}(\rho_{\rm d}(\mu)|J|)^3 k_{\rm B}T
\frac{\partial^2}{\partial T^2}\left( C_1TK(t)+C_2 TI(t) \right),
\end{equation}
where we defined
\begin{equation}
I(t)= {\rm Im}\psi'\left( \frac{1}{2}+i\frac{\mu}{\pi T}\right).
\end{equation}
Thus the log-$T$ terms do not show up in the region $T\sim |\mu|$.

In the high-temperature region defined by $|\mu|\ll T$, 
the temperature dependence of $V$ comes from the terms of order $\mu/T$.
Thus the additional entropy is also of order $\mu/T$ and is 
negligible in the region $|\mu|\ll T$.

\section{Summary}

We investigated the Kondo problem with the localized spin which couples to
Dirac fermions.
The Kondo temperature $T_{\rm K}$ was calculated from a singularity of Green's functions.
The logarithmic terms in the resistivity and the specific heat were derived in
the low-temperature region, where
The region $k_{\rm B}T\ll |\mu|$ is called the low-temperature region.
We considered two regions:\\
\\
(1) When $k_{\rm B}T\ll |\mu|$, the Kondo screening occurs
with the characteristic temperature scale
\begin{equation}
k_{\rm B} T_{\rm K} \simeq \bar{D}\exp\left( -\frac{8}{\rho_{\rm d}(\mu)|J|} \right),
\end{equation}
for small $\rho_{\rm d}(\mu)|J|\ll 1$.
This is the conventional form, as in the original Kondo problem, being proportional
to $\exp(-{\rm const}/\rho |J|)$ with the density 
of states $\rho$.
When $|\mu|$ is small compared to the cutoff $D$, $T_{\rm K}$ is proportional
to $|\mu|$:
\begin{equation}
k_{\rm B} T_{\rm K} \simeq |\mu|\exp\left( -\frac{8}{\rho_{\rm d}(\mu)|J|} \right).
\end{equation}
In the range $T_{\rm K} < T\ll |\mu|/k_B$, the log-$T$ anomaly appears in
the physical quantities.
For large $\rho_{\rm d}|J|$ of order 1, $T_{\rm K}$ will be given by an
algebraic function $Q(x)$ such as 
\begin{equation}
k_{\rm B}T_{\rm K}\simeq |\mu|Q(\rho_{\rm d}(\mu)|J|). 
\end{equation}
\\
(2) When $k_{\rm B}T$ is not much smaller than $|\mu|$, namely, $k_{\rm B}T\sim |\mu|$
or $k_{\rm B}T>|\mu|$, the Kondo effect is suppressed and
the resistivity and specific heat do not exhibit a logarithmic (log-$T$) anomaly.
$T_{\rm K}$ vanishes when the chemical potential is just at the Dirac point,
leading to the absence of Kondo screening.

The vanishing of $T_{\rm K}$ when $\mu$ is at the Dirac point is consistent with
the results for the pseudogap Kondo problem\cite{voj06}, and suggests the
existence of a phase transition, as $|J|$ is increased, from non-screening 
phase to screening phase.
The term being proportional to $\sigma_z$ like the magnetic field 
would also be important in the Kondo effect.
which we did not consider in this paper.
It is also interesting to study the nature of interaction between two magnetic 
impurities in Dirac metals.  This issue was investigated intensively in
the original Kondo problem\cite{jay81, jon89, yan91, aff95}. 
These are future subjects.

The author expresses his sincere thanks S.A. Jafari, I. Hase and K. Yamaji 
for discussions.
He also thanks M. Vojta, E. Orignac and A. Principi for useful comments.

\end{document}